# Identity-based Trusted Authentication in Wireless Sensor Network

Yusnani Mohd Yussoff[1], Habibah Hashim[2] and Mohd Dani Baba[3]

[1,2,3] Computer Engineering Department, University Teknologi MARA,
Shah Alam, Selangor 40450, Malaysia
**yusna233;habib350;mdani074@salam.uitm.edu.my**

**Abstract**

Secure communication mechanisms in Wireless Sensor Networks (WSNs) have been widely deployed to ensure confidentiality, authenticity and integrity of the nodes and data. Recently many WSNs applications rely on trusted communication to ensure large user acceptance. Indeed, the trusted relationship thus far can only be achieved through Trust Management System (TMS) or by adding external security chip on the WSN platform. In this study an alternative mechanism is proposed to accomplish trusted communication between sensors based on the principles defined by Trusted Computing Group (TCG). The results of other related study have also been analyzed to validate and support our findings. Finally the proposed trusted mechanism is evaluated for the potential application on resource constraint devices by quantifying their power consumption on selected major processes. The result proved the proposed scheme can establish trust in WSN with less computation and communication and most importantly eliminating the need for neighboring evaluation for TMS or relying on external security chip.

Keywords: *Trusted, Security, Authentication, Wireless Sensor Network, Identity-based cryptography*

## 1. Introduction

Wireless Sensor Networks (WSNs) is network consisting of sensor nodes or motes communicating wirelessly with each other. Advancement in sensor, low power processor, and wireless communication technology has greatly contributed to the tremendous wide spread use of WSNs applications in contemporary living. Example of these applications include environmental monitoring, disaster handling, traffic control and various ubiquitous convergence applications and services[1]. Low cost and without the need of cabling are two key motivations towards future WSN applications. These applications however demand for considerations on security issues especially those regarding nodes authentications, data integrity and confidentiality. Commonly, the sensor nodes are left unattended, and are vulnerable to intruders. The situation becomes critical when the nodes are equipped with cryptographic materials such as keys and other important data in the sensor nodes. Moreover, adversaries can introduce fake nodes similar to the nodes available in the network which further leave the sensor nodes as untrusted entities. Two approaches have been widely researched to ensure the validity of nodes in the networks thus further confirm the need of trusted communication between nodes in the network. The following paragraph briefly discusses the two approaches.

TMS is one of the more widely used mechanisms in aiding WSN member (trustors) in dealing with the uncertainties in participants (trustees) future actions [2]. It basically studies the behaviour of the nodes in the networks for a certain period and calculates the trust value. However, TMS can only detect the existence of fake nodes in the network after a certain period. Hence, adversary nodes may have participated in the network and may have caused network disorders by the time TMS identifies them. Furthermore, since TMS is mathematical-based it indirectly imposes burdens to sensor nodes such as extra processing power, memory requirement, and communication in the networks.

The node's trustworthiness can also be achieved through the Trusted Platform Module (TPM) crypto-processor chip. In a recent work, Wen Hu [3, 4] used the TPM hardware which is based on Public Key (PK) platform to augment the security of the sensor nodes. It was claimed that the SecFleck architecture proposed in [3] provides the internet-level PK services with reasonable energy consumption and financial overhead. Unfortunately, the drawbacks of TPM chip which include extra hardware entailed on the platform and the superfluous of commands required to perform its functions both contribute to higher energy utilizations.

To avoid the infeasibility of deploying TPM chip in wireless sensor nodes, this study proposes the use of the ARM1176JZF-S processor with Trustzone features as described in [5].

This paper proposes a secure mechanism to accomplish a trusted relationship between sensors in the wireless networks according to TCG specifications. Firstly it describes how the trusted platform is established; follows by the description on trusted authentication protocol that confirms only trusted nodes existed in the network. Finally it presents an analysis on the energy consumption for the trusted platform and the authentication protocol.

The remainder of this paper is structured into six major parts: Section 2 addresses the current security challenges in WSNs followed with some introductory notes on trust as outlined by TCG and Identity Based Encryption (IBE) in section 3 and 4 respectively. Section 5 introduces the design for a trusted platform based on ARM1176JZF-S processor. Further, Section 5 describes the proposed IBE-Trust security framework. Then the analysis on the proposed scheme is discussed in Section 6. Finally, Section 7 concludes the paper.

## 2. Security in WSNs

Security mechanisms for WSNs can be divided into three related phases. The first phase is to secure the sensor node or the platform itself so that the network originator can guarantee the integrity of the sensor node of the network. Next phase is the big challenge in securing the network infrastructure or the wireless medium to ensure reliable, secure and trusted communication. The final phase involves protecting the confidentiality and integrity of the data since in wireless communication anyone can intercept the data. Hence these three components namely the sensor node, network infrastructure and data are the crucial entities that need to be protected in wireless sensor network. This served as the fundamental requirement in the design of a trusted wireless sensor framework. The following sub-sections present the proposed security goals and the simplified TCG specifications adopted as the basis in the design of secured framework.

### 2.1 Proposed Security Goals

In acknowledging the various types of attacks in WSNs as discussed in [6], the secured framework in this study proposes the following security features.

*Trusted Platform* - Trusted Platform is achieved through a chain-of-trust with image identified as "bootloader1" in the SoC ROM as the Root of Trust (ROT) and a secure boot process that measures the integrity of software images, applications, and components on the sensor nodes. Also, the trusted platform offers secure memory location for sensitive credentials such as private keys.

*Trusted Authentication* - Verifies that a sender is a trusted user or node and will behave in a trusted manner for the network. The authentication protocol which is developed on Identity based Cryptography is identified as IBE_Trust. This protocol confirms the authenticity of nodes and also the confidentiality and integrity of the exchanged message.

### 2.2 TCG Specifications for Trust Establishment

According to [7], trust can be defined as an entity that always behaves in an expected way for any intended functions. The basic properties of a trusted computer or system can be listed as follows;

- *Isolation of programs* – prevents program A from accessing data of program B
- *Clear separation between user and supervisor process* – there should be a system to prevent user applications from being interfered by the operating system.
- *Long term protected storage* – secret values are stored in a place that last across power cycles and other events.
- *Identification of current configuration* – provides identity of the platform as well as software or hardware executing on it.
- *Verifiable report of the platform identity and current configuration* – a way for other users to validate a platform.
- *Include Hardware based protection*- protection in a combination of hardware and software.

The basic building block of a trusted platform according to TCG definition consists of *properties*, *measurement* and *reporting*. Properties refer to unique or unaltered values over the life of the platform. Measurement is the process of obtaining the identity of the platform function and should begin at the ROT of the platform. It will measure the hash value of the platform component before passing the control to the next process. The flow of the measurement process is called the 'Chain – of –Trust'. The ROT is an entity that must be trusted as well as properly protected as there is no mechanism available to measure it. Finally, the reporting will provide the evidence to those wishing to rely on the information and is established through report or attestation.

Note that trust is established through two different processes which are measurement and reporting or attestation. In order to ensure message integrity and confidentiality during the reporting process, the message will be encrypted using Identity Based Encryption (IBE) algorithm. Brief discussion on IBE is presented in the subsequent section.

## 3. Identity-Based Encryption (IBE)

IBE was proposed by Adi Shamir in 1984 and only in 2001, Boneh and Franklin [8] have successfully implemented a fully functioning IBE scheme. The IBE has simplified the certificate based public key encryption scheme by using publicly known unique identifiers to derive public keys and eliminate the needs of certificate authority.

In IBE, an arbitrary string is used as a public key. The public key can be calculated from any string such as email, project name or any other string. According to RFC 5408 [9], an IBE public key can be calculated by anyone who has the essential public key while a cryptographic secret (master key) is needed to calculate the IBE private key, in which the calculation can only be performed by a trusted server that has this secret. In WSN, the trusted authority or trusted entities is the BS which has to be placed in the most secured place and controlled directly by the network proprietor. Besides that, the existence of pre-deployment stage offers better security and controlled environment for the key distribution phase. This criterion does not exist in other Public Key Cryptography (PKC) infrastructure.

Another characteristic that differentiates IBE from other server-based cryptography is that no communication is required with the server during encryption operation whereby the sender only needs to know the recipient's ID to encrypt the message. Additionally, IBE implementation also consumes less memory for storing public keys of the other nodes. These factors have supported the use of IBC instead of PKC in this implementation. Fig. 1 portrays the difference in concept between PKC and IBC followed by four stages in standard IBC implementation.

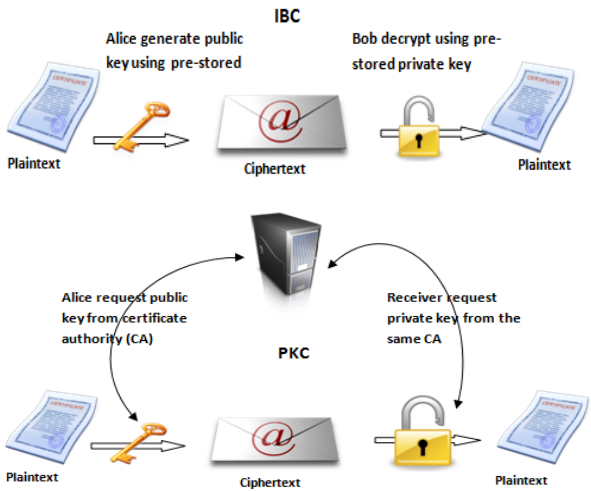

Fig. 1 IBC and PKC standard implementation

*Setup* – This process should be done by any Trusted Agent (TA). In WSNs, TA can be the BS. A security parameter $k$ is provided as the input and BS will generates the public BF parameters ($G_1$, $G_T$, $ê$, $n$, $P$, $sP$, $H_1$, $H_2$, $H_3$, $H_4$) and its master key, $s$. The parameters are pre-loaded to all sensor nodes in the network. Interested readers can refer to the book by Luther Martin [10] for more details.

*Extract* - The extract process needs public parameters and master key values from the setup process. The public keys associated with sensor node ID are identified by mapping the identity on the elliptic curve $E/F_q$: $y^2 = x^3 + 1$ using Eq. (1). The outcome from the cryptographic hash function $Q_{IDx}$ is then multiplied with the master key, $s$ to obtain the private key $d_x$.

$$Q_{IDx}=H_1(ID_x) \qquad (1)$$
$$d_x=sQ_{IDx} \qquad (2)$$

**Encrypt** – The input to this process includes common parameters, recipient ID and message M $\in$ M and the output ciphertext C $\in$ C.

$$C = encrypt(params,ID,M) \qquad (3)$$

**Decrypt** – The input to this process are common parameters, private key $d_x$, and C $\in$ C while the output is M $\in$ M.

$$M=decrypt(params,d_x,C) \qquad (4)$$

## 4. Framework of Trusted Sensor Node

This section discusses the methods used to accomplish the previously mentioned security features. It is divided into two major sections which are identified as *Trusted Platform* and *IBE-Trust* for simplicity.

### 4.1 Trusted Platform

The security provided by cryptography mainly depends on safeguarding the cryptographic keys from adversaries. It grants the need to adequately protect the keys to ensure confidentiality and integrity of sensitive data. This section discusses on how this study manipulates ARM1176JZF-S security features to fulfil the TCG trust definition for a trusted platform. Listed are ARM1176JZF-S features that are used to realize the basic properties of trusted platform design. Fig. 2 correlates TCG trust specifications with the proposed solution.

*Secure world* – sensitive resources such as encryption and decryption images will be placed in the secured world memory locations. Trust Zone Address space controller

(TZASC) is used to configure regions as either secure or non-secure. All non-secure processes will be rejected from the secure region. This ensures the confidentiality of important data and images.

*Single physical core* – safe and efficient execution of code from both normal and secure world. Secure monitor codes are developed to switch from normal to secure and vice versa.

*On-SoC RAM and ROM* - will ensure no highly sensitive data leaves the chip thus reduce the possibility of physical attacks.

*Secure boot* – a process to ensure the integrity of the software images and devices on the platform and generate management value as platform unique entity.

*IBE-Trust protocol* – confirming secure communication between sensors and BS.

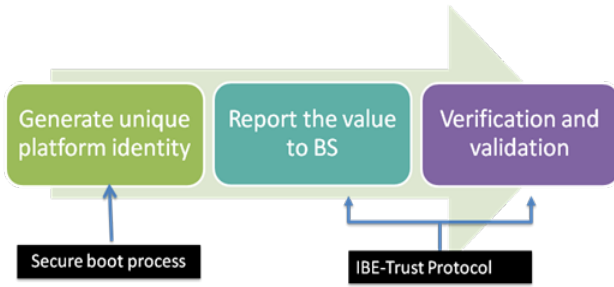

Fig. 2 Process flow of the framework

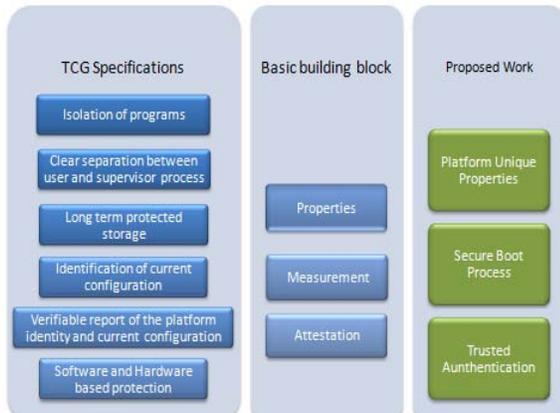

Fig. 3 Correlation between TCG specification and our proposed solution

### 4.2 Secure Boot Process as the Chain-Of-Trust

The overall process was developed based on ARM1176JZF-S development board. Codes are written in assembly language to minimize memory size and to speed up processing time. At the point of writting, this study has successfully developed a secure boot process up to Level 2 ($L_2$) of a total of three levels. The proposed *Chain-of-Trust* is best described as follows:

**Level 1 ($L_1$):** The ROT which is the entity that must be trusted is located in the 16KB on-SoC ROM of the ARM1176JZF-S processor. The integrity ($I_0$) of the image that is burned into it which is the $1^{st}$ Boot loader image is assumed to be unmodifiable and therefore is always TRUE. This assigns 1 to level 1 ($L_1$)

ROT $\rightarrow$ Boot Loader ($BL_1$)[assume trust] $\rightarrow$ Integrity ($I_1$) = True= 1

$\therefore L_1 = 1$

**Level 2 ($L_2$):** Verifies image of the second bootloader ($BL_2$) residing in the external storage by measuring the hash value of the image. The referenced value is predetermined and is stored together with the first level bootloader. If the integrity of $2^{nd}$ bootloader ($I_2$) is verified, then the $2^{nd}$ bootloader image is loaded and executed.

Hash ($BL_2$)' == Hash value of [($BL_2$) in $BL_1$] $\rightarrow$ Integrity ($I_2$) = True = 1

$\therefore L_2 = 1$

Where the prime symbol " ' " substitutes as the new measure values.

At this stage, if $I_2$ equals to 0, the process will halt. The sensor node will be able to complete the secure boot process only if the integrity in each level is true. Once successful, the unique value generated from the secured boot process will then be used to establish the trust relationship with the BS. Due to the limited register space in ARM1176JZF-S, the secured boot design will only consider eight hexadecimal characters as the comparison value. For validation, hundreds of different images were hashed using SHA-2 algorithm and it was found that none of the output produced an identical eight hash value in a location. For security reason, the location is undisclosed. The secure boot integrity (I) is checked using the Boolean equation as in Eq. (5).

$$I = I_1.I_2.I_3......I_{N-1}.I_N \qquad (5)$$

Where, *N* represents the level in the secured boot process or the last entity in the chain of trust. The integrity checking is transitive from 1 to 2 to 3 and to N and does not invert where trusting entity 1 does not imply to trusting entity N and trusting entity N requires trusting entity 1 to N-1.

### 4.3 SHA-256

SHA is a type of cryptographic hash function that guarantees the integrity. As the security-performance tradeoff is relatively linear, two factors are identified contributing to the selection of the hash algorithm. First is the size of the algorithm itself; it must be small enough to fit into the secured location and the second is that the algorithm must be powerful enough to resist from attacks with no collision [11] in the algorithm. This study takes advantages of the 256-bit SHA-2 as the hash algorithm. Although there are several algorithms in the SHA family, SHA-2 has proven to be safe in literatures to date [12]. SHA-2 output 64 hexadecimal characters or 256-bit hash values and is considered as "sufficiently high" for the foreseeable future. Moreover the run time necessary for a birthday attack is on the order of $2^{128}$ and therefore is currently assumed to be collision free.

### 4.4 Secure and Non-secure world

Trustzone architecture of the ARM processor enables the construction of a programmable environment that allows the confidentiality and integrity of almost any asset to be protected from specific attacks. In other words, there are two different modes in ARM processor. Normal mode allows access to all system resources while secure mode restricts access to resources. Trustzone state is controlled by the Secure Monitor Code (SMC) that handles switching between secured and non-secured world. SMC requires complex codes to allow calls from complex Real Time Operating System (RTOS). Other method is by specifying the secured and non-secured region in the scatter file. Sensitive processes such as SHA-2, encryption and decryption were configured to run in a secured environment by calling the monitor switch function prior to process. This is straight forward and sufficient for currently proposed system.

### 4.5 Address Space Partitioning and Interrupts

Trustzone address spaces are divided into secure (only accessible in trust world) and non-secure regions (accessible from both state). TrustZone Protection Controller (TZPC) is one of the ways used to configure different regions in the memory as secured or non-secured. However, this work defines secured and non-secure regions using page-table file because it was found much simpler and less complex. The world in which the processor is executed is indicated by the non-secure bit (NS-bit) in the secured configuration register (CP15). Low value of NS-bit indicates the secured world execution. IRQ and FIQ are two interrupt vectors that are used to switch the processor into monitor mode.

The abovementioned sections have discussed the methods to accomplish the first two basic building block of becoming a trusted platform which are property and measurement. Following sections discuss on communication procedures in registering valid sensor nodes into the network utilizing the unique entity derived earlier, thus fulfill the third specification which is reporting through attestation or report.

## 5. IBE-Trust Security Model

Typical WSNs scenarios adopted in the proposed framework are uncontrolled environment, random node placement and self configuration. The networks consist of several sensor nodes and a BS as the trusted agent. All sensor nodes communicate via bi-directional wireless link with equal transmission range. Each node has a unique, string based, non-zero identity and are loosely synchronized. During the first implementation, all sensors in the network will report its ID to BS.

Standard four IBE stages have been reduced to three stages in the implementation. The earlier two stages which are setup and extract are combined together. The combination of the two stages was made possible due to the proposed IBE implementation procedure. The overall development used the Tate Pairing algorithm by [13] downloaded from Shamus website [14]. The MIRACL library was than compiled into ARM single image library and was included in the executable images (ibe_gen, encrypt, and decrypt) to benchmark elliptic curve point manipulation.

For implementation, this study suggests four different stages starting from generation of keys and common parameters to on-line node registration. Scopes of the different stages are discussed in the following subsections.

### 5.1 Delivery phase (DP)

The DP stage is offline with the intention to provide the networks with complete information such as the identity of sensor nodes, private keys, master key, and BF parameters except master key. All newly joined nodes need to go through this stage thus allowing the BS to have a list of nodes appear in the network.

### 5.2 Pre-Deployment (PDP)

Once configured with the necessary information, the sensor nodes will go through a boot-up process under a controlled environment to generate its unique management value. The generated value together with the sensor node ID will securely be sent to BS for further verification.

### 5.3 Deployment Stage (DY)

This process happens immediately after node deployment at the intended location. At this stage the sensor node will boot up and go through the secure boot process. Outcome from this stage is the same unique trust value.

### 5.4 Trusted Authentication (TA)

This stage aims to register node's unique ID with BS for further communication. Successful boot up node will report it trust value to the trusted authority, which in this case is the BS. The BS will then decrypt the message, verify the unique ID together with the trust value in its database. Upon successful authentication, BS will generate a new list containing the trusted node's identity (trustID). This new list, which is smaller than the trust list will be distributed to sensors in its network for faster verification process between nodes. To this stage, this study has not finalized any secure distribution methods of trustID table to existing nodes in the network.

### 5.5 Packet format

According to IEEE 802.15.4 compliant radio transceiver standard, the maximum packet length is 127 bytes. However, maximum data payload for CC2420 transceiver according to TinyOS packet format is 114 bytes. However, to enable extra information for IBE_trust protocol, the maximum data size or payload is now reduced to 106 bytes. The IBE_trust packet consists of 2 bytes sender ID, 2 bytes random nonce value, bytes message and 4 bytes truncated MAC. Fig. 4 depicts the packet format starting with raw message followed by payload data structure according to TinyOS and finally IBE_trust packet format.

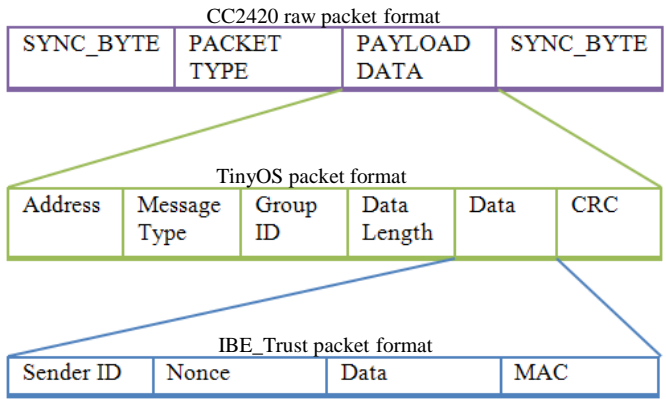

Fig. 4 Packet Structure

Trusted nodes in the network remain in trusted condition as long as it remains in the ON state. Once rebooted or shutdown for any reason, the nodes will need to re-authenticate with the BS. Failure to authenticate will lead to node termination process where the node's ID will be removed from the trust list. Formal analysis of the protocol will be discussed in our later publications.

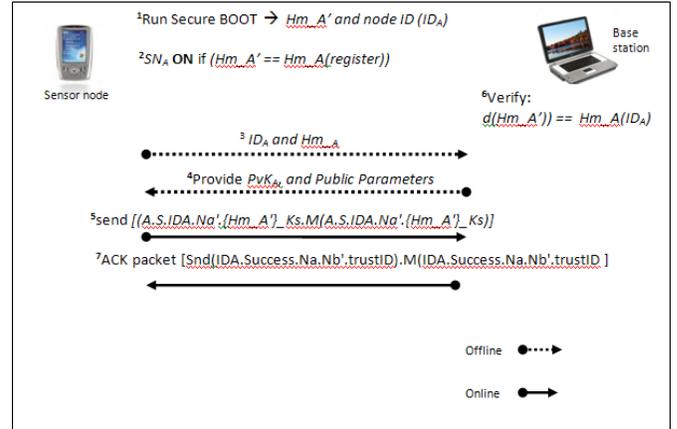

Fig. 5 IBE_Trust Authentication protocol

### 5.6 ID based-one-pass Authenticated Key Exchange (AKE)

Due to availability of trust list in each sensor, subsequent communications between sensors are very much simplified. Receiving sensor will be authenticated based on sender ID and upon successful, receiver will locally generate session key using pre-installed key derivation function (KDF) based on the receiving value for its subsequent secure communication. To utilize the installed parameters, the AKE is based on the symmetric bilinear pairings.

A: Picks random number $r \in Z^*_q$,

Computes $R = rQ_A$ where $Q_A$ is public key of A and send R to B over public channel using packet format as depicted in Fig. 5.

A → B: $Snd(A.B.ID_A.R'.Na'.Mac(ID_A.R'.Na'))$

where $h = H_2(R,ID_A\|ID_B)$ and is computed by both parties and $S_A$ is the private key of node A which is securely stored in On-SoC ROM. Both parties A and B then compute the shared secret as $K_{AB} = e((r+h)S_A, Q_B)$ and $K_{BA} = e(R + hQ_A, S_B)$ and finally the session key is computed by A as $\mathcal{K}(K_{AB})$ and by B as $\mathcal{K}(K_{BA})$ where $\mathcal{K}$ is key derivation function.

The idea towards this implementation is adopted from ID-based one-pass AKE technique [15] and the only difference is in the authentication value where, [28] authenticated using sender public key and this work used sender ID.

Suggested applications for the proposed scheme include health and medical monitoring where nodes assigned to users may first have to register with the BS and once successful they are free to move and data can travel securely direct to the BS or in multi-hop manner.

## 5.7 Energy Equation

To confirm the practicability of the proposed IBE-trust scheme in WSNs, this study also calculates the energy consumption for a newly joined sensor node. Due to the availability of switching process between secured and non-secured modes in ARM1176JZF-S, the switching energy is added to the energy consumption equation for more accurate values. The total energy for sending encrypted data (unique trust value) to the BS is calculated as in Eq. (6). All energy values are presented in joules using Eq. (9).

$$E_T = E_{Boot} + E_{SW} + E_{enc/bit} * data(bits) + E_{ta} \quad (6)$$

$$E_{ta} = E_{Tx} * bytes\ transmitted + E_{Rx} * bytes\ received \quad (7)$$

Substituting (7) into (6), makes (8):

$$E_T = E_{Boot} + E_{SW} + E_{enc} * data(bits) + E_{Tx} * bytes + E_{Rx} * bytes \quad (8)$$

$$E(J) = Power\ in\ watts * time \quad (9)$$

## 5.8 Notations

Table 1 describes the notations used in the proposed scheme.

Table 1: Notations used in the proposed scheme

| Symbol | Description |
|---|---|
| $ID_A$ | Identifier of sensor node A |
| $PvK_A$ | Private key of sensor node A |
| A.S | Sender ID.Receiver ID |
| Hm_A' | New trust value (DY) stage |
| Hm_A | Trust value at PDP stage |
| $N_a'$, $N_b'$ | Random Nonce |
| $K_s$, $K_A$ | Public key BS and A |
| Snd | Send packet |
| _Ks | Encrypted packet with $K_S$ public key |
| Mac | Hash function |
| M | Message |
| $S_{id}$, $R_{id}$ | Sender ID and Receiver ID |
| d(Hm_A) | Decrypt(Hm_A) |
| ACK | Acknowledgement packet |
| $E_{Boot}$ | Secure boot-up energy |
| $E_{SW}$ | Energy in Switching process |
| $E_{Tx}$, $E_{Rx}$ | Transmit and Receive energy |
| $E_{enc/bit}$ | Encryption energy per bit |
| $E_{ta}$ | Trusted Authentication energy |

## 6. Analysis of the proposed scheme

This section presents the analysis of the proposed scheme.

### 6.1 Energy Utilization

The results are obtained by conducting the analysis on the ARM1176JZF-S development board. The processor runs at 20mA, 3.6V with frequency 667MHz. Since the encryption and communication processes consume most of the energy[16], this study only considers the amount of energy used by these processes. As part of the benchmarking, this study also compares the work with secFleck [3] implementation that utilizes TPM chip in providing the public key technology for WSN.

The energy per bit used in the encryption process for secFleck (hardware/software) is 5.4/7030μJ while in this study the energy/bit is 22.5 μJ used for the encryption process which is fully implemented using software. Although the energy used in the proposed scheme is higher compared to SecFleck hardware based implementation, it does not employ external crypto-processor chip on the sensor node platform.

Based on a preliminary testing in this study, the switching process takes about 0.23s and consumes around 16.56mJ of energy which is higher than that required for the encryption process. This limits switching from normal to secure mode and vice versa for important processes only. However the delay can be reduced in actual implementation where function calls to clock and standard input-output can be eliminated.

Tate pairing as seen in Table 3 consumed the highest energy due to its complicated computation. However, this study obtains 0.148J lower than the result obtained by Doyle et al. [17] that utilized ARM7. This shows an indirect relationship between the processor specifications and sensor node lifetime. Hence, it implicates the use of dedicated low power processor for embedded applications. To realize computational complexity of trusted authentication and authenticated key exchange, energy utilization of the above processes is calculated and presented. As energy to transmit and receive is proportional to message size, total energy is calculated using equation 6. Assuming 106 bytes payload and 21 bytes header, nodes performing trusted authentication process needs to transmit an encrypted message of 280 bytes consisting of key file and cipher text message and receive an acknowledgement packet with list of trusted node ID (one time only during node deployment).

Assuming 200 nodes, size of trustID will be around 400bytes and total bytes received are around 480 bytes including packet header ((400/106)*127). For nodes to nodes authentication and key exchange, node only needs to send a single message, sized 85 bytes (64 bytes of $rQ_A$ and 21 bytes header) to its neighbor. Table 4 tabulated the energy consumption based on CC2420 transceiver used in the proposed work.

Table 2: Energy consumption for major processes

| Process | Delay(s) | Energy |
|---|---|---|
| Secure Bootup (1st stage only) | 0.059 | 4.24mJ |
| Encryption (C++) | 0.05 | 22.5µJ/bit |
| Sha2 (asm) | 0.05 | 3.6mJ |
| Switching (asm+ C) | 0.23 | 16.56mJ |
| Fast tate pairing | 4.05s | 0.292J |

Table 3: Communication overhead of our trusted authentication scheme based on CC2420 transceiver

| Proposed Work | Process | Bytes (data + header) | Energy |
|---|---|---|---|
| Trusted Authentication | Transmit (1.83µJ/byte) | 319 | 0.58mJ |
| | Receive (1.98 µJ/byte) | 480 | 0.95mJ |
| Key exchange | Transmit (1.83µJ/byte) | 85 | 0.15mJ |
| | Receive (1.98 µJ/byte) | 0 | 0 |

Total energy used for a one time trusted authentication process calculated using equation (8) is 0.027J. Energy incurs during Fast tate pairing is not included as its can be done offline. Assuming nodes with limited 1000J full battery capacity [17], the percentage of energy used for the above processes is to be less than 1%. It is believed that the results obtained are an acceptable cost for one-time or rare distribution of trust management values to establish trust relationship between sensor nodes and the BS.

## 6.2 Efficiency

To confirm the efficiency of the proposed scheme, the comparison of energy utilization for user authentication scheme is tabulated in Table 4. Data for existing work in the table were adopted from Rehana's et. al [18] work. It is clearly seen that the proposed mechanism consumes the least energy as compared to other schemes for the same security features of user authentication and secure communications.

Table 4: Energy comparison of proposed user authentication scheme with proposed scheme

| Schemes | Authentication scheme | Energy Costs (mJ) | Storage Overhead (bytes) | Session Key |
|---|---|---|---|---|
| RRUAN | ECDSA | 106.84 | 0 | No |
| DP²AC | RSA | 14.05 + TE | 10N | No |
| Rehana[18] | IBS | 72.90 | 0 | Yes |
| Proposed Scheme | IBE-trust+one-way AKE | 26.9 | 2N | Yes |

** N = number of nodes

## 6.3 Security Analysis

This section generally demonstrates how the proposed protocol can prevent typical attacks on sensor networks.

### 6.3.1 Physical Attacks

The use of ARM1176JZF-S as the processor with its On-Soc memory has helped in this study to protect important credentials such as sensor node private keys. Moreover, in this scheme, only part of the private key is stored in the sensor node memory thus further protect the sensor nodes and network. Images such as encryption and decryption are stored in the secured memory region of flash memory and are only accessible in the secured mode environment. The effect of BSL attacks can also be reduced through the secure boot process where the integrity of loaded images has been verified to prevent sensor nodes from running malicious code.

### 6.3.2 Node impersonation

Node impersonation happens when intruders manage to duplicate the unique identity of the sensor node that is being used during authentication. Non-regeneration of the same trust value through secure boot process has significantly reduced the possibility of having a masquerade node in the network.

### 6.3.3 Typical wireless attacks

This study also confirms that the communication during trusted authentication is free from active attack such as message modification, replay attack, false message through packet encryption, nonce value as well as entity and data authentication. The confirmation is done through formal analysis method and is discuss in another paper.

### 6.3.4 Security of the proposed scheme

The security of IBE-Trust protocol is best realized through the security of full BF IBE scheme. In this scheme, the public key can be written as $Q_{ID} = tP$ for some unknown t. Therefore $ê(rQ_{ID},sP) = ê(rtP,sP) = ê(P,P)^{rst}$ and ciphertext $C = (rP, M \oplus H_2ê(P,P)^{rst})$. If an adversary manages to get $P$ and $sP$ from the public parameters, they can calculate $Q_{ID} = tP$ from receiver's identity and observes $rP$ in the ciphertext. Moreover, if the adversary manages to calculate $ê(P,P)^{rst}$ from $P, sP, rP$ and $tP$ then it will be able to recover the message M by calculating $(M \oplus H_2ê(P,P)^{rst} \oplus H_2(ê(P,P)^{rst})) = M$. Calculating $ê(P,P)^{rst}$ is actually solving Bilinear Diffie-Hellman Problem (BDHP) and is very difficult [10]. This brief analysis confirms the confidentiality of unique trust value of each sensor node in the network that is sent to BS encrypted with IBE scheme.

For node to node authentication, our proposed scheme uses the ID-based one-pass AKE. In the existing protocol presented by [15], the authentication is established when both parties manage to generate similar shared key locally. In our proposed protocol, beneficiary node will first check the identity of nodes requesting to authenticate and proceed to compute the secret shared key if the ID exists in the table provided by BS. This somehow has provided a two tier security mechanism and has limited this expensive operation to valid nodes only. Identity-based one-pass AKE is based on symmetric bilinear pairings and is secure by assuming the hardness of BDHP with $H_1, H_2$ and $K$ modeled as random oracle. Interested readers can find proof to this method in [15].

## 7. Conclusion

This paper has presented an alternative method to confirm the trustworthiness of nodes in WSN. The proposed scheme involves designing a trusted platform and an energy efficient authentication protocol. For the trusted platform, ARM1176JZF-S processor together with CC2420 chip has been chosen as the platform processor and transceiver respectively. Both processor and transceiver chips have greatly supported the design of low energy trusted platform. Besides low energy, most importantly the proposed trusted platform fulfills the trust requirement as outlined in the TCG documentation. Consequently, the proposed trusted mechanism has contributes to enhance security in WSNs by reducing the probability of fake or clone sensor node through non-regenerated unique platform identity. Finally, the proposed work has opened a new research area towards trusted sensor node platform.

## Acknowledgments

The authors would like to thank Research Management Institute, Universiti Teknologi MARA for financial support and members of WSN and Trusted Computing team for their ideas and helpful feedback.

**Yusnani Mohd Yussoff** graduated with B.Sc (Hons) in Electrical Electronic Engineering majoring in Computer from University Science Malaysia and currently a PhD candidate of Universiti Teknologi MARA, Shah Alam, Malaysia. Current research area is in Wireless Sensor Network focusing on trusted wireless sensor node platform.

**Second Author** Habibah Hashim received her PhD from Universiti Tenaga Nasional, Malaysia in 2007. She obtained her BSc (Hons) in Electrical and Electronics Engineering from University of Nottingham, UK and her MSc in Computer Aided Engineering from Teesside University, UK. Currently she is a Deputy Dean of Research and Industrial Linkages and Associate Professor at the Faculty of Electrical Engineering, Universiti Teknologi MARA. Her research interests are Computer Networks, Wireless Networks, Trusted Computing and Informatics.

**Mohd Dani Baba** received his PhD from University of Sussex, UK in 1996. Currently he is a Professor in Computer and Communication Engineering at the Faculty of Electrical Engineering, Universiti Teknologi MARA. His research interests are computer networking, cloud computing, wireless networking and mobile communication.